# XHIP-II: Clusters and associations

Charles Francis[1], Erik Anderson[2]
[1] 25 Elphinstone Rd., Hastings, TN34 2EG, UK.
[2] 800 Morton St., Ashland, OR 97520, USA.

*Context*. In the absence of complete kinematic data it has not previously been possible to furnish accurate lists of member stars for all moving groups. There has been an unresolved dispute concerning the apparent inconsistency of the Hipparcos parallax distance to the Pleiades.
*Aims*. To find improved candidate lists for clusters and associations represented among Hipparcos stars, to establish distances, and to cast light on the Pleiades distance anomaly.
*Methods*. We use a six dimensional fitting procedure to identify candidates, and plot CMDs for 20 of the nearest groups. We calculate the mean parallax distance for all groups.
*Results*. We identify lists of candidates and calculated parallax distances for 42 clusters and 45 associations represented within the Hipparcos catalogue. We find agreement between parallax distance and photometric distances for the most important clusters. For single stars in the Pleiades we find mean parallax distance 125.6 ± 4.2 pc and photometric distance 132 ± 3 pc calibrated to nearby groups of similar in age and composition. This gives no reason to doubt either the Hipparcos database or stellar evolutionary theory.

## 1. Introduction

Complete and accurate memberships of kinematic groups are important for the study of Galactic dynamics, for tests of stellar evolutionary theory, and to estimate cluster distances, required to establish the cosmological distance scale. However, the absence of complete kinematic information and inevitable uncertainties in stellar parallaxes create difficulties in ascertaining the validity of lists of candidate stars for individual clusters. Furthermore, it is now established that the local velocity distribution contains six clearly defined stellar streams (e.g., Famaey et al., 2005; Francis and Anderson, 2009, 2012 and references cited therein). In consequence, accidental alignments between proper motions and positions are far more probable than would be the case for a well mixed distribution.

We define moving groups as stars sharing a common motion and localized in a region of space. They are distinguished from streams, which are all-sky motions and which are parts of the spiral structure of the Galaxy (Francis & Anderson, 2009, 2012). Moving groups will be termed 'clusters' if they are gravitationally bound, and 'associations' otherwise. Associations typically consist of young stars originating in the same process, resulting from the collisions between outward bound gas clouds (corresponding to the Hyades stream) and clouds following the spiral arm. In this paper a separation parameter will be defined, and used to determine whether a group is likely to be a cluster or an association.

Because of the likelihood of chance alignments it is important to calculate group memberships on the basis of complete kinematic data, and with the most accurate and extensive information available. *The Extended Hipparcos Compilation* ("XHIP", Anderson & Francis, 2012) is based on *Hipparcos the New Reduction of the Raw Data* ("HIP2", van Leeuwen, 2007) gives radial velocities for 46 392 Hipparcos stars, together with metallicities for 18 549 stars and multiplicity information from the *Catalog of Components of Double & Multiple Stars* (Dommanget & Nys, 2002) and *The Washington Visual Double Star Catalog*, version 2010-11-21 (Mason et al., 2001-2010). It is therefore appropriate that XHIP recalculates group memberships for all clusters and associations containing a reasonable number of Hipparcos stars with known radial velocities.

We tested a number of groups using membership lists taken from the literature, and all groups listed in the *New catalog of optically visible open clusters and candidates (V3.0)* (Dias et al., 2002-2010) or in Karchenko et al. (2005a&b) with distances less than 800 pc in either catalogue, at which point the groups have become too distant to contain many Hipparcos stars, and parallax distances have become meaningless. We used a six dimensional (position × velocity) fitting procedure which results in improved accuracy for group memberships as indicated by a larger numbers of candidates and/or smaller dispersion than found in lists obtained from the literature. We found evidence in Hipparcos stars for 87 clusters and associations, but we note that it does not necessarily follow from lack of evidence that a group does not exist, since it may contain less bright stars. 42 groups show sufficient separation from the surrounding star field to be classed as probable clusters.

The distance to the Pleiades has been the subject of an unresolved dispute. Van Leeuwen (2009) has suggested that the difference between the Hipparcos and other estimates is due to a failure of stellar evolutionary theory, while others have questioned the Hipparcos astrometry. Van Leeuwen calculated the Pleiades parallax distance using a weighted mean. We observe that weighting can lead a systematic error. After recalculating without weighting we find a distance within 2σ error bounds of most other estimates and conclude that this gives no reason question either stellar evolutionary theory or the Hipparcos astrometry. We also resolve anomalous results concerning the distance of the Diamond cluster and Blanco 1, and, because of the importance of cluster distances to the cosmological distance scale, we recalculate all parallax distances without weighting.

Section 2. reviews the algorithm by which we determine cluster memberships in XHIP. In section 3. we describe the method used to determine cluster distance. Section 4. describes the colour-magni-tude diagrams (CMDs) for a number of significant groups. Section 5. contains remarks on particular groups. Section 6. comments on the Pleiades distance anomaly, finding that the most likely cause for any remaining anomaly is random error. Our conclusions are summarised in section 7..

## 2. Group membership

The traditional method for ascertaining membership of the Hyades has been the convergent point method (e.g. Perryman et al., 1998). The convergent point method does not make use of parallax distance, which has to be added as a separate constraint. With the accuracy of parallaxes now available for stars at the distance of the Hyades it is no longer the best method (Van Leeuwen, 2009) for the Hyades. For groups at greater distance, with lower proper motions and (importantly) lower radial velocities, errors dominate and the convergent point method is unreliable. Van Leeuwen's approach was to first find groups in the celestial sphere (RA and DE coordinates) and then to constrain the other variables. In consequence, group membership is determined by a cylindroid in position × velocity space, rather than the more natural choice of an ellipsoid described below, and because the initial selection depends on only two variables it is difficult to accurately determine cluster size so that his lists typically omit candidates which we find at the edges of a cluster. We also have a database with many more, and more accurate, radial velocities than previous studies, and it makes sense to use this additional information in finding improved candidate lists.

To identify the membership of a moving group it is necessary to match both the position and the velocity of each star to within a region of a six-dimensional space, where distance is determined from Hipparcos parallax measurements. We have rejected the notion that membership can be decided from other factors, such as stellar composition, because indicators of composition, such as [Fe/H] are notoriously uncertain (as is seen, for example, in XHIP, in which measurements of [Fe/H] from many sources were compared and combined).

The list of group members is determined iteratively, by testing each star in XHIP under the condition:

$$\sum_x \frac{(x-\bar{x})^2}{\sigma_x^2} < n^2 \tag{2.1}$$

for a fixed value of $n$, where $x$ runs over the dynamical variables, *Plx*, *RA*, *DE*, *RV*, *pmRA*, *pmDE*, except for groups occupying a large region of the sky, when $x$ runs over the Cartesian variables $X$, $Y$, $Z$, $U$, $V$, $W$, where $X$ is directed towards the Galactic centre, $Y$ is the direction of rotation, $Z$ is directed to the Galactic north pole, and $U$, $V$, $W$ are the respective velocity components. $\bar{x}$ and $\sigma_x$ are the mean and standard deviation for the group satisfying eq. (2.1) in the previous iteration. It is necessary to use Cartesian variables when the angular size of the cluster substantially affects radial velocity and proper motions, but it is better to use radial variables where possible because they correspond more closely to the actual measurements and because radial distance errors are much greater than angular errors, and hence the error ellipsoid is aligned with the radial direction. Cartesian variables were used for the Hyades cluster, the Ursa Major association, and the AB Doradus, β Pictoris and Tucana/Horologium moving groups.

At each iteration, all stars in XHIP with known *RV* that satisfy eq. (2.1) are included in the new group list. Then the means and variances of the variables $x$ for the group are recalculated and the iteration is repeated. The convergence criterion is simply that the candidate list remains unchanged in an iteration.

We initialized the iteration using $n = 3.5$ and a list of group stars taken from the literature, when available, or group coordinates given in Dias et al. or Karchenko et al., together with typical group dimensions. The Hyades was initialised from the list given by Perryman et al. (1998). The β Pictoris, AB Doradus, Tucana/Horologium, TW Hydrae associations and the η Chamaeleontis cluster were initialised from Zuckerman & Song (2004). Upper Centaurus Lupus, Lower Centaurus Crux, Upper Scorpius, Vel OB2, Collinder 121, Per OB2, Cas-Tau, Lac OB1, Cep OB2, and Cep OB6 were initialised from de Zeeuw et al. (1999). The cluster was considered "found" when the group became stable under iteration of eq. (2.1). For associations, the found groups are not necessarily those intended by those who originally named them. For example, ASCC 18 and ASCC 20 converge to the same list of stars. For a number of associations, convergence was found for larger groups than those originally intended. Our lists contain overlaps. For example, the Orion Molecular Cloud Complex contains a number of groups including the Orion Nebula (M42), the Horsehead Nebula, Collinder 70, NGC 1980, NGC 1981, and the Running Man Nebula (NGC 1977) which have similar motions and which can be regarded as parts of the larger group. In the case of an overlap more than one comma separated group is listed in column *GrpName* of XHIP.

If an initial list containing at least four Hipparcos stars with known radial velocities was not found from the data in any of our sources we rejected the cluster. Otherwise we found a revised list by iterating with eq. (2.1). Usually the size of the candidate list increased or stabilised at each iteration, but for some groups the number of candidates reduced before stabilising. If the number of candidates continued to increase without stabilising, or if the dispersion of any of the dynamical variables was unusually large for a group we used $n = 3$, which generally resulted in a more compact list of candidates. We then tried $n = 3.5$ again. If a stable group was found under the iteration we increased $n$ by 0.5 and repeated the iteration to establish the largest half integral value of $n$ for which a stable group under the iteration could be found.

For most well-known clusters our 6-dimensional fitting procedure leads to a larger number of candidates and/or smaller dispersion than is found in lists obtained from the literature. For groups using spherical coordinates, we extended the list of candidate stars to include stars without known radial velocities by reducing the right-hand side of eq. (2.1) so as to leave the standard deviation of each dynamical variable similar to its value for the core group of stars with complete dynamical information. This method does not apply to groups found using Cartesian coordinates.

Eq. (2.1) can be justified both dynamically and statistically. The dynamical argument has intuitive appeal, but the statistical argument may be preferred because it incorporates both measurement errors and physical dispersion, does not depend on approximation, and applies to unbound groups. Dynamically, for a gravitationally bound group, the further a star is from the centre of the group the nearer its velocity must be to that of the centre of gravity of the cluster. This can be described using a linear approximation for the gravitational potential energy at distance $r$ from the centre of the cluster, such that $r^2 + kv^2 \approx$ constant, where $v$ is velocity relative to the centre of the cluster, and the constant $k$ is determined by the ranges of values of $r$ and $v$ taken for each star over time. Since $r$ and $v$ are bounded for cluster stars, we may impose eq. (2.1) as a criterion for cluster membership. Statistically eq. (2.1) defines the interior of an ellipsoid in position × velocity space, centred at the mean for the cluster, and with axes proportional to the standard deviation for each variable. Thus our procedure generalises to 6D the familiar principle of discarding outliers beyond a given number of standard deviations from the mean.

Group identifications can only be given with certainty for stars with known radial velocities, and accurate parallaxes. In practice, many stars still do not have known radial velocities and, even using HIP2, parallax errors lead to distance errors much greater than the size of the group. Consequently, there remains some uncertainty in group memberships. Associations are not gravitationally bound, and do not have the ellipsoidal shape suggested by eq. (2.1). Memberships of associations are necessarily less accurate than those of clusters, but eq. (2.1) still has merit, by providing an objective criterion by which the existence of a group and a list of candidate stars can be established.

$n$ is a measure of the concentration of a group, and of its separation from the surrounding star field, and (because it is rigorously defined) may be preferred to the Trumpler classifications I - IV. Since they are gravitationally bound, clusters are more compact than associations. As a result the largest value of $n$ giving convergence can be expected to be greater for a cluster than for an association. In practice, we found that a maximum value of $n = 3.5$ is typical for an association. For clusters $n \geq 5$ is usual. The lower value of $n$ for associations shows poor dynamical separation from surrounding stars, indicating that the majority of associations are just randomly dense regions arising in much greater processes in which stars are formed. This was also shown by the fact that a number of associations found in the Pleiades stream have overlapping memberships. We rejected groups with $n < 3.5$ or with fewer than four Hipparcos stars with known radial velocities.

Seven groups have separation $n = 4.5$: Feigelson 1, α Perseus, Cep OB6, ASCC 122, the Christmas Tree cluster, ASCC 115, and ASCC 113. These are all young groups, and can be classed as OB associations. However, both α Perseus and the Christmas Tree cluster are generally regarded as clusters, and have 1σ radii less than 10 pc. It seems probable that a separation of $n = 4.5$ is indicative of a gravitationally bound group, with less good separation from the surrounding star field than one would expect of a mature cluster, due to the similar motions of other stars in the complex in which the group formed. We have therefore classed 42 groups with $n \geq 4.5$ as clusters, and 45 groups with $n \leq 4$ as associations.

## 3. Parallax distance

Van Leeuwen (2009) calculated group parallax distances using weighted mean parallax. Weighting by inverse squared error is generally a sound method to combine different measurements and reduce uncertainty, but it requires that there is no correllation between the values and weights. We confirmed that there is no correlation between parallax and parallax error for the entire population (there is no reason why there should be), but the method used by Hipparcos compares parallax to the surrounding star field. Although van Leeuwen has rejected this as a cause of systematic

**Clusters and associations within 300 pc**

| Cluster | $N$ | $N_{vL}$ | Sep $n$ | $R_{XHIP}$ pc | $R_{vL}$ pc | $R_D$ pc | $R_K$ pc | 1σ rad pc | 1σ rad deg |
|---|---|---|---|---|---|---|---|---|---|
| Ursa Major | 15 | 8 | 6 | 25.2 ± 0.3 | 25.3 | 25 | | 3.2 | 7.3 |
| β Pictoris | 43 | | 3.5 | 27.8 ± 2.1 | | | | 24.6 | 118 |
| AB Doradus | 28 | | 3.5 | 29.3 ± 2.4 | | | | 22.1 | 131 |
| Hyades | 190 | 150 | 5.5 | 46.1 ± 0.6 | 46.4 ± 0.5 | 45 | | 8.1 | 10.1 |
| Tucana / Horologium | 73 | | 3.5 | 50.1 ± 1.2 | | | | 48.1 | 73.8 |
| Coma | 43 | 27 | 6 | 86.3 ± 1.3 | 86.7 ± 0.9 | 96 | 87 | 7.2 | 4.8 |
| Feigelson 1 | 7 | | 4.5 | 110.8 ± 7.0 | | 114 | | 3.9 | 2.0 |
| Lower Centaurus Crux | 187 | | 3.5 | 122.4 ± 1.4 | | | | 19.1 | 9.0 |
| Pleiades | 70 | 53 | 6 | 124.1 ± 2.7 | 120.2 ± 1.9 | 133 | 130 | 5.8 | 2.7 |
| Upper Centaurus Lupus | 169 | | 3.5 | 143.6 ± 2.0 | | | | 23.3 | 9.3 |
| Upper Scorpius | 82 | | 3.5 | 144.0 ± 2.7 | | | | 8.7 | 3.5 |
| Southern Pleiades (IC 2602) | 29 | 15 | 6 | 148.8 ± 3.2 | 148.6 ± 2.0 | 161 | 160 | 3.4 | 1.3 |
| Omicron Velorum (IC 2391) | 20 | 11 | 5.5 | 165.6 ± 8.0 | 144.9 ± 2.7 | 175 | 176 | 2.7 | 0.9 |
| Platais 2 | 7 | | 4 | 178 ± 22 | | 201 | 201 | 8.2 | 2.6 |
| Platais 9 | 7 | | 8 | 178.9 ± 7.5 | | 174 | 200 | 6.9 | 2.2 |
| Praesepe | 40 | 24 | 6 | 181.8 ± 8.5 | 181.5 ± 6.0 | 187 | 187 | 7.3 | 2.3 |
| NGC 2451A | 20 | 14 | 6.5 | 186.0 ± 4.8 | 183.5 ± 3.7 | 189 | 188 | 6.0 | 1.8 |
| α Perseus | 80 | 50 | 4.5 | 186.6 ± 4.4 | 172.4 ± 2.7 | 185 | 190 | 9.6 | 2.9 |
| Blanco 1 | 16 | 13 | 6 | 205 ± 15 | 207 ± 12 | 269 | 269 | 3.6 | 1.0 |
| Collinder 135 | 14 | | 5 | 247.3 ± 5.6 | | 316 | 319 | 8.4 | 2.0 |
| NGC 1647 | 6 | | 6 | 248 ± 38 | | 540 | 539 | 4.4 | 1.0 |
| Stephenson 1 | 15 | | 4 | 265 ± 16 | | 390 | 373 | 4.5 | 1.0 |
| Ptolemy (NGC 6475) | 22 | 20 | 5.5 | 266 ± 11 | 270 | 301 | 300 | 3.3 | 0.7 |
| IC 348 | 28 | | 3.5 | 274 ± 16 | | 385 | 394 | 9.9 | 2.1 |
| Cep OB6 | 11 | | 4.5 | 274.8 ± 5.7 | | | | 14.9 | 3.1 |
| Platais 4 | 30 | | 3.5 | 280 ± 18 | | 276 | 276 | 16.2 | 3.3 |
| NGC 7092 | 14 | 7 | 6 | 292 ± 15 | 303 | 326 | 311 | 4.7 | 0.9 |
| Stock 2 | 12 | | 4 | 294 ± 28 | | 303 | 380 | 6.2 | 1.2 |
| Collinder 359 | 10 | | 4 | 299 ± 27 | | 249 | 640 | 8.7 | 1.7 |

**Table 1:** The number of members, $N$, and cluster distance, $R_{XHIP}$, are given for the full candidate lists, including multiple stars and stars with unknown radial velocity. $N_{vL}$ and $R_{vL}$ are as given by van Leeuwen (2009). $R_D$ and $R_K$ are distances given by Dias et al. and Karchenko et al. Sep is the separation parameter, $n$, which describes the concentration of the group and its kinematic separation from field stars.

**Clusters and associations beyond 300 pc**

| Cluster | $N$ | $N_{vL}$ | Sep $n$ | $R_{XHIP}$ pc | $R_{vL}$ pc | $R_D$ pc | $R_K$ pc | 1σ rad pc | 1σ rad deg |
|---|---|---|---|---|---|---|---|---|---|
| NGC 2232 | 10 | 6 | 6 | 320.2 ± 9.7 | 352.1 | 359 | 325 | 4.6 | 0.8 |
| NGC 6405 | 7 | | 5.5 | 322 ± 32 | | 487 | 487 | 3.6 | 0.6 |
| Orion Molecular Cloud | 231 | | 3.5 | 366.6 ± 9.4 | | | | 32.0 | 5.0 |
| ASCC 18/20 | 29 | | 3.5 | 322 ± 18 | | 450 | 450 | 8.8 | 1.6 |
| ASCC 21 | 30 | | 3.5 | 324 ± 18 | | 500 | 500 | 9.4 | 1.7 |
| ASCC 19 | 34 | | 3.5 | 328 ± 18 | | 350 | 350 | 11.0 | 1.9 |
| NGC 2451B | 19 | | 3.5 | 331 ± 18 | | 302 | 430 | 11.9 | 2.1 |
| Collinder 69 | 9 | | 5.5 | 343 ± 25 | | 400 | 438 | 10.0 | 1.7 |
| vdBergh-Hagen 23 | 12 | | 3.5 | 356 ± 30 | | | 437 | 3.3 | 0.5 |
| Collinder 70 | 179 | | 3.5 | 368 ± 10 | | 387 | 391 | 23.6 | 3.7 |
| Trumpler 10 | 5 | 12 | 5.5 | 369 ± 34 | 387.5 | 424 | 417 | 5.6 | 0.9 |
| ASCC 16 | 11 | | 3.5 | 376 ± 16 | | | 460 | 8.3 | 1.3 |
| Collinder 350 | 7 | | 6 | 380 ± 28 | | 280 | 280 | 11.0 | 1.7 |
| Collinder 140 | 12 | 9 | 6 | 388 ± 42 | 376.0 | 405 | 402 | 4.4 | 0.6 |
| Diamond cluster (NGC 2516) | 21 | 11 | 5 | 388 ± 22 | 342.5 | 409 | 346 | 6.1 | 0.9 |
| Vel OB2A | 14 | | 4 | 389 ± 23 | | | 411 | 6.0 | 0.9 |
| Briceno 1 | 14 | | 4 | 395 ± 16 | | 460 | | 8.6 | 1.3 |
| NGC 2422 | 8 | 7 | 6 | 399 ± 38 | 396.8 | 490 | 491 | 2.0 | 0.3 |
| IC 4665 | 9 | 7 | 6 | 407 ± 68 | 355.8 | 352 | 352 | 3.0 | 0.4 |
| NGC 2547 | 25 | 8 | 3.5 | 413 ± 21 | 474.0 | 361 | 457 | 7.9 | 1.1 |
| NGC 6882 | 8 | | 5 | 419 ± 57 | | | 342 | 11.8 | 1.6 |
| NGC 2547 core | 5 | | 9.5 | 422 ± 49 | | 361 | 457 | 0.9 | 0.1 |
| NGC 3532 | 9 | 6 | 5.5 | 423 ± 59 | 415.5 | 486 | 497 | 4.0 | 0.5 |
| Pozzo 1 | 32 | | 3.5 | 425 ± 23 | | 336 | | 11.4 | 1.5 |
| Alessi 19 | 4 | | 5 | 441 ± 58 | | 550 | 550 | 6.2 | 0.8 |
| IC 4756 | 7 | | 5.5 | 445 ± 105 | | 484 | 484 | 4.4 | 0.6 |
| Sigma Orionis | 15 | | 3.5 | 446 ± 30 | | 399 | 399 | 9.1 | 1.2 |
| ASCC 127 | 6 | | 4 | 449 ± 23 | | 350 | 350 | 5.3 | 0.7 |
| BH 99 | 20 | | 4 | 450 ± 38 | | 507 | | 17.0 | 2.2 |
| ASCC 122 | 38 | | 4.5 | 454 ± 31 | | 700 | 700 | 15.1 | 1.9 |
| Stock 1 | 7 | | 5.5 | 455 ± 60 | | 318 | 350 | 6.3 | 0.8 |
| Christmas tree | 9 | | 4.5 | 470 ± 91 | | 667 | 660 | 4.9 | 0.6 |
| Lac OB1 | 67 | | 3.5 | 466 ± 16 | | | | 34.5 | 4.2 |
| BH 23 | 24 | | 3.5 | 479 ± 49 | | 480 | | 9.2 | 1.1 |
| ASCC 100 | 16 | | 4 | 486 ± 90 | | 350 | 350 | 10.0 | 1.2 |
| NGC 1981 | 12 | | 4 | 487 ± 29 | | 400 | 400 | 7.2 | 0.8 |
| Vel OB2B | 73 | | 3.5 | 510 ± 18 | | | 411 | 25.8 | 2.9 |
| Running Man | 29 | | 3.5 | 513 ± 28 | | 500 | 500 | 13.8 | 1.5 |
| Roslund 5 | 18 | | 3.5 | 516 ± 76 | | 389 | 418 | 12.0 | 1.3 |
| Orion Nebula | 30 | | 3.5 | 519 ± 30 | | 414 | 399 | 11.9 | 1.3 |
| Collinder 132 | 8 | | 5.5 | 536 ± 34 | | 472 | 411 | 11.9 | 1.3 |
| M 41 (NGC 2287) | 8 | | 13 | 545 ± 92 | | 710 | 693 | 1.0 | 0.1 |
| Stock 7 | 4 | | 6 | 550 ± 105 | | | 698 | 0.6 | 0.1 |
| Alessi 12 | 9 | | 4 | 588 ± 42 | | 537 | 537 | 13.3 | 1.3 |
| Stock 23 | 6 | | 4 | 633 ± 115 | | 380 | 380 | 4.5 | 0.4 |
| NGC 1980 | 30 | | 3.5 | 638 ± 90 | | 550 | 550 | 12.8 | 1.1 |
| NGC 7160 | 8 | | 3.5 | 668 ± 96 | | | 794 | 8.8 | 0.8 |
| NGC 6639 | 6 | | 3.5 | 705 ± 209 | | 700 | 700 | 12.8 | 1.0 |
| ASCC 115 | 6 | | 4.5 | 778 ± 99 | | 600 | 600 | 12.1 | 0.9 |
| Kronberger 28 | 5 | | 5 | 785 ± 148 | | 550 | | 2.8 | 0.2 |
| Cep OB3 | 9 | | 3.5 | 788 ± 114 | | | 700 | 11.7 | 0.9 |
| ASCC 113 | 8 | | 4.5 | 801 ± 124 | | 450 | 450 | 11.0 | 0.8 |
| ASCC 29 | 15 | | 3.5 | 807 ± 132 | | 750 | 750 | 18.0 | 1.3 |
| IC 2395 | 9 | | 3.5 | 818 ± 109 | | | 709 | 8.1 | 0.6 |
| Cep OB2 | 13 | | 3.5 | 883 ± 85 | | | | 23.7 | 1.5 |
| NGC 2548 | 5 | | 16 | 926 ± 664 | | | 769 | 3.0 | 0.2 |
| IC 4725 | 4 | | 9 | 1955 ± 2149 | | 620 | 620 | 4.7 | 0.1 |
| Ruprecht 89 | 17 | | 3.5 | 2505 ± 909 | | 750 | 750 | 44.8 | 1.0 |

**Table 2:** Columns as in table 1.

error in stellar parallaxes, he did not exclude the possibility of an effect on parallax errors. We tested the correlation between parallax and parallax error for forty clusters (omitting two with overlaps between groups) using a two tailed Student's *t*-test. We found that eight contain correlations significant at 95% or more, and three significant at 99% or more. For the Diamond cluster (one of those for which van Leeuwen reported a distance anomaly) the correlation between parallax and parallax error is -0.6 and is significant at 99.95%. In the case of the Pleiades the correlation is -0.2 and is significant at 86%. Both positive and negative correlations are found. A detailed analysis of the reasons for these correlations would be complex and have little value, but the existence of such correlations means that an unweighted mean must be used when calculating cluster distance.

For more distant stars, van Leeuwen derived cluster parameters from a combined analysis of the abscissa residuals for the member stars. This method uses intermediate astrometric data for single cluster members. Because of the importance of avoiding weighting, in all cases we have calculated distance from the mean parallax. This is valid even for distant clusters where some stated parallaxes are negative. It results in larger stated errors than a method using a weighted mean, but we believe this is correct, so as to avoid unknown systematic errors.

In practice, in many cases the distance estimate is little changed by the use of weighting in the calculation of mean parallax, but the result of using weighting is notably low for the Pleiades, Omicron Velorum (IC 2391) and α Perseus (table 1), and for the Diamond cluster (NGC 2516), IC 4665 (table 2). In three cases, NGC 2232, Trumpler 10 and NGC 2547, for which van Leeuwen found greater distances, the difference is largely accounted for by changes in the candidate lists.

The full list of clusters and properties is given in groups.dat, which is part of XHIP, available as CDS Catalog V/137B. groups.dat is described in table 3. Distances given in groups.dat are based on the full candidate list, including stars for which radial velocities are unknown. In practice we found small differences, well within error bounds, when these stars were excluded. For (nearby) groups whose membership is calculated using Cartesian variables Dist is mean distance, otherwise Dist is found from mean parallax. In table 1 and table 2, $R_{\mathrm{XHIP}}$ is Dist, given in groups.dat.

Because of difficulties in the solution from the raw data, parallaxes are inevitably less accurate for multiple star systems than for single stars. In most cases restricting to single stars makes little difference to results, but it is worth remarking that for the group of three clusters for which anomalously low parallax distances were found by van Leeuwen, the Pleiades, Blanco 1 and the Diamond cluster (NGC 2516), restricting to single stars caused an increase in the distance estimate, in the case of Blanco 1 by as much as 5%.

| Label | Units | Explanation |
|---|---|---|
| Name | --- | Group name |
| AltName | --- | Alternative group name |
| Count | --- | Number of Hiipparcos candidate stars |
| n | --- | Separation parameter (27) |
| Stream | --- | Kinematic group (MNRAS ref) |
| RAdeg | deg | Group mean right ascension (ICRS) |
| DEdeg | deg | Group mean declination (ICRS) |
| Dist | pc | Group distance (28) |
| pmRA | mas/yr | Group mean proper motion in RA*cos(DEdeg) |
| pmDE | mas/yr | Mean proper motion in Declination |
| RV | km/s | Group mean radial velocity |
| s_RAdeg | deg | Group width on RA*cos(DEdeg) (1sigma) |
| s_DEdeg | deg | Group width on DE (1sigma) |
| s_Dist | pc | Group width on Dist (1sigma) |
| s_pmRA | mas/yr | Group width on pmRA (1sigma) |
| s_pmDE | mas/yr | Group width on pmDE (1sigma) |
| s_RV | km/s | Group width on RV (1sigma) |
| e_RAdeg | deg | Standard error on RA*cos(DEdeg) |
| e_DEdeg | deg | Standard error on DE |
| e_Dist | pc | Standard error on Dist |
| e_pmRA | mas/yr | Standard error on pmRA |
| e_pmDE | mas/yr | Standard error on pmDE |
| e_RV | km/s | Standard error on RV |
| rad | pc | Physical radius (1sigma) |
| arad | deg | Angular radius (1sigma) |
| X | pc | Heliocentric distance towards Gal. center |
| Y | pc | Heliocentric distance towards Gal. rotation |
| Z | pc | Heliocentric distance towards N. Gal. Pole |
| U | km/s | Heliocentric velocity towards Gal. center |
| V | km/s | Heliocentric velocity towards Gal. rotation |
| W | km/s | Heliocentric velocity towards N. Gal. Pole |
| s_X | pc | Group width on X (1sigma) |
| s_Y | pc | Group width on Y (1sigma) |
| s_Z | pc | Group width on Z (1sigma) |
| s_U | km/s | Group width on U (1sigma) |
| s_V | km/s | Group width on X (1sigma) |
| s_W | km/s | Group width on X (1sigma) |
| e_X | pc | Standard error on X |
| e_Y | pc | Standard error on Y |
| e_Z | pc | Standard error on Z |
| e_U | km/s | Standard error on U |
| e_V | km/s | Standard error on V |
| e_W | km/s | Standard error on W |
| [Fe/H] | [Sun] | ? mean Iron abundance |
| e_[Fe/H] | [Sun] | ? Standard error on [Fe/H] |
| n_[Fe/H] | --- | Number of Hipparcos member stars with [Fe/H] |
| E(B-V)_K | mag | ? Color excess in B-V, CDS Cat. J/A+A/477/165 |
| E(B-V)_D | mag | ? Color excess in B-V, CDS Cat. B/ocl |

**Table 3:** Description of groups.dat, contained in CDS Catalog V/137B.

## 4. Colour-magnitude diagrams

We plotted the CMD for three nearby moving groups, AB Doradus, β Pictoris, and Tucana/Horologium, together with the Ursa Major association, the Hyades, the Coma star cluster, Praesepe, the Pleiades, the Southern Pleiades and Omicon Velorum (figure 1). The motions of AB Doradus, β Pictoris, and Tucana/Horologium, ensure that they consist of young stars. We plotted only single stars, since they have more accurate parallaxes and more accurate luminosities, leading to a tighter distribution. For the Pleiades we used an applied reddening $E(B-V) = 0.034$ mag for $RA > 056.818968^{\circ}$, and $E(B-V) = 0.057$ mag otherwise (Taylor, 2008). For other clusters $E(B-V)$ was taken from Karchenko et al. We used extinction equal to 3.1 times reddening (e.g. Gibson & Nordsieck, 2008).

The tight form of the CMD in figure 1 is a measure of the accuracy of Hipparcos parallaxes for near stars and clusters. In particular there is good agreement between the positions of β Pictoris and Tucana/Horologium and AB Doradus. The small difference between the position of these groups and the Hyades is accounted for by the greater age and metallicity of the Hyades. The Pleiades does appear slightly below and to the left of the other groups. This could in part be caused by an underestimate of the effect of extinction, but is most probably the result of random errors (section 6.).

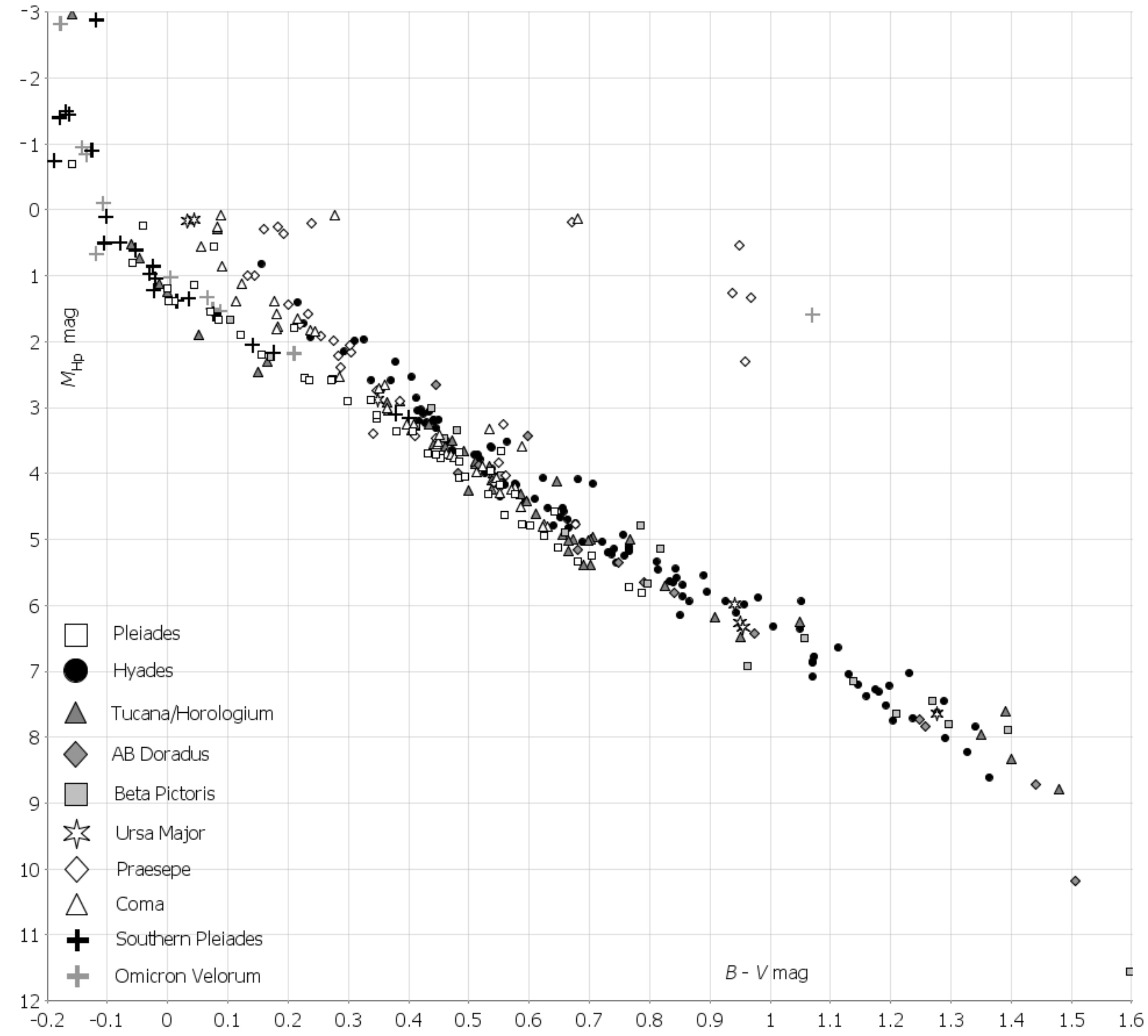

**Figure 1:** CMD for single stars in the Pleiades cluster, the Hyades cluster, the β Pictoris, AB Doradus and Tucana/Horologium moving groups, the Ursa Major association, the Coma star cluster, Praesepe, the Southern Pleiades and Omicron Velorum. Cluster distance is used for the Pleiades, Coma, Praesepe, Southern Pleiades and Omicron Velorum, and parallax distance for individual stars is used for the other groups. Reddening has been applied.

It is possible to read the relative ages of particular groups from the positions of the main sequence cutoff and the main sequence for early types. Thus the Hyades appears to be a little older than Praesepe, not coeval as is sometimes assumed (e.g. Dobbie et al., 2006). Praesepe and the Hyades are both members of the Hyades stream, formerly known as the Hyades supercluster, and have similar motions, but it is known that streams contain stars with a wide range of ages (e.g., Chereul et al., 1998, 1999). Praesepe does appear coeval with the Coma cluster, but is clearly not related as they have very different motions. Figure 1 confirms that these clusters are older than the Pleiades, AB Doradus, β Pictoris, and Tucana/Horologium.

We plotted the CMD for single stars in the Hyades & Pleiades clusters, NGC 2451A, α Perseus, Blanco 1, Collinder 135, the Ptolemy cluster, NGC 7092, NGC 2232, Collinder 140, the diamond cluster and the Scorpio-Centaurus association (figure 2). With the exception of the Hyades and Pleiades, which are included in both CMDs for comparison, most of these groups are younger than those in figure 1, as may be seen from the difference in the curve followed by stars on the main sequence. In accordance with the predictions of evolutionary theory, early types are below and to the left of the corresponding position in figure 1, while for colours in the range ~0.4 < $B - V$ < ~0.8 mag we find pre-main sequence stars lying above the main sequence. Greater scatter is also seen in figure 2 due to the spread of distances of stars in the Scorpius-Centaurus association.

## 5. Notable groups

### *5.1. Ursa Major association*

The Ursa Major association (Collinder 285) has a core containing 15 Hipparcos stars with complete kinematic information and $n$ = 6. Two stars, HIP 65327 and HIP 59496 are added to the nucleus given by King & Villarreal (2003) and by Soderblom & Mayor (1993). The value of $n$ is exceptionally high for an association, showing low velocity dispersion and good separation from surrounding stars. This supports the notion that the Ursa Major association is the remnant of an open cluster. There are no further candidates in Hipparcos for which radial velocities are not available. No convergence is found from extended lists of Ursa Major candidates (e.g. Soderblom & Mayor, 1993; King & Villarreal, 2003). We believe that such extended lists contain stars with similar velocities taken from the Sirius stream, and should not be regarded as belonging to the Ursa Major association.

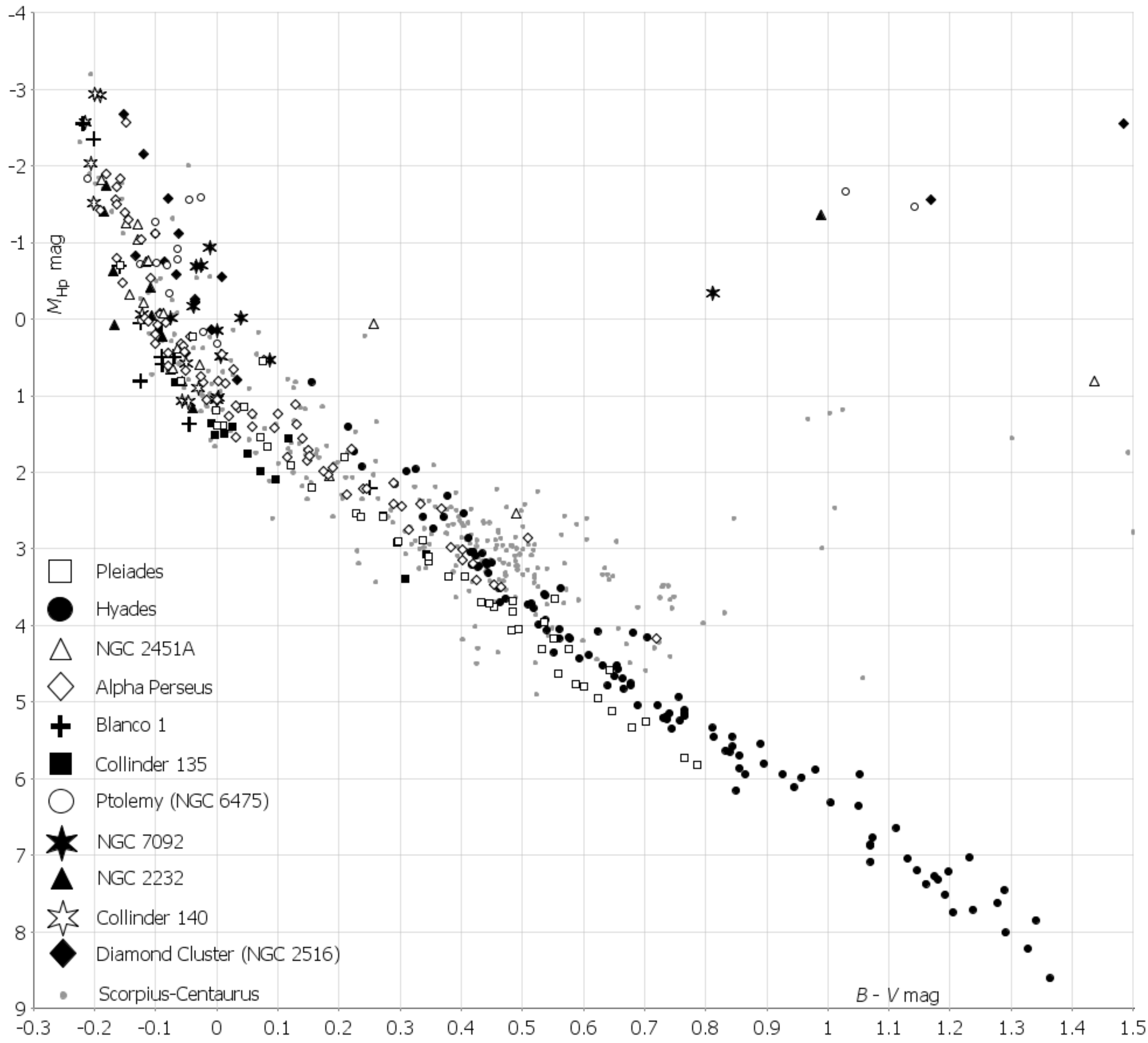

**Figure 2:** CMD for single stars in the Hyades & Pleiades clusters, NGC 2451A, α Perseus, Blanco 1, Collinder 135, the Ptolemy cluster, NGC 7092, NGC 2232, Collinder 140, the diamond cluster and the Scorpio-Centaurus association. Parallax distance for individual stars is used for the Hyades and cluster distance is used for the other groups. Scorpio-Centaurus uses distances for Lower Centaurus Crux, Upper Centaurus Lupus and Upper Scorpius. Reddening has been applied. There is greater scatter than in figure 1 because of the spread of distances in the Scorpius-Centaurus association, and because of pre-main sequence stars lying above the main sequence, as well as very young stars lying below it.

### *5.2. Hyades cluster*

The Hyades (Melotte 25) are sufficiently close that inaccuracies due to change in angular position must be excluded when calculating group membership. Errors in distance are modest and membership is best calculated using the dynamical variables $X$, $Y$, $Z$, $U$, $V$, and $W$. We restricted to stars with better than 20% parallax errors. We identified 190 Hipparcos stars in the Hyades cluster with known radial velocities, and with cluster separation $n$ = 5.5. From the list of 197 Hyades candidates given by Perryman et al. (1998), eight do not have known radial velocity, two more have parallax errors greater than 20%, and 14 are rejected as outside our dynamical cluster definition, leaving 173 in our list. Six dimensional fitting resulted in more compact distributions in position and velocity space.

We may expect our estimate of the distance to the Hyades to be slightly low because the constraint that parallax errors are less than 20%, which is necessary to calculate corrected motions in *UVW* space, introduces a selection bias toward nearer stars. As only two stars were rejected by this criterion, the bias is small. In fact our distance to the centroid of the cluster is 45.5 ± 0.8 pc, a little less than the mean distance 46.1 ± 0.6 pc (as one expects), slightly below other estimates but well within a 1σ error.

### 5.3. Pleiades stream

Velocities found for Lower Centaurus Crux, Upper Centaurus Lupus, Tucana/Horologium, and β Pictoris are very similar, and close to the modal velocity of the Pleiades stream, which consists (in the main) of very young stars. Many other stars have similar velocities. Similarly, AB Doradus has a similar velocity to the Pleiades cluster and numerous associations with similar motions take part in the Orion Molecular Cloud Complex. This supports the model of spiral structure described by Francis & Anderson (2009, 2012) in which star forming regions are formed from collisions between outward bound gas clouds and inward bound gas following the spiral arm. These collision processes take place continuously and on a large scale. Associations are best regarded as denser regions of a much greater process, such that associations formed near to each other in space and time are likely to have similar velocities and composition.

### 5.4. OB associations

De Zeeuw et al. (1999) found membership lists for a number of OB associations, without benefit of known radial velocities, using a variant on the convergent point method. The convergent point method works well for the Hyades, which is at relatively low distance, has relatively high proper motion, and, importantly, has high radial velocity. However, this method is unreliable for associations at greater distances such that errors are large compared to proper motion, and such that the convergent point is sufficiently far from the cluster that non-linear affects due to the curvature of the celestial sphere come into play.

The Scorpius-Centaurus association contains three sub-associations, Lower Centaurus Crux, Upper Centaurus Lupus, and Upper Scorpius. The list for Upper Scorpius from de Zeeuw et al. contains 120 candidates, over a region with 1σ radius 13.7 pc., and with radial velocity dispersion 8.9 km $s^{-1}$, high for stars which are supposed to be kinematically related. Our fitting procedure restricted the 1σ radius to 8.7 pc and radial velocity dispersion to 4.7 km $s^{-1}$. Dispersion in proper motion was also around half of that of the original list. We found 82 candidates in a region with less than 4% of the original 6-volume. Our final list for Upper Centaurus Lupus contains 169 members, compared to 221 in the list from de Zeeuw et al., in a region of less than 7% of the 6-volume. Our list for Lower Centaurus Crux contains 187 members, compared to 180 in the list from de Zeeuw et al., in a region less than 18% of the size.

After initialising Vel OB2 using coordinates from Karchenko et al., we found 14 candidates, including 9 with known radial velocities, with separation n = 4 and distance 389 ± 85 pc, centered at *RA* =122.2º, *DE* = – 48.9º, and with radial velocity 14.8 ± 1.9 km $s^{-1}$. After initialising from the list of 93 candidates for Vel OB2 given by de Zeeuw et al. we found 73 candidates, in a region of 5% of the size. Only 39 of our candidates are in common with those of de Zeeuw et al.'s list. Only 28 of our candidates have known radial velocities. We found a distance of 510 ± 18 pc, at *RA* = 122.3º, *DE* = – 46.7º, radial velocity 22.2 ± 1.0 km $s^{-1}$. Only three candidates are in common with our list for the nearer group, and we conclude that these are separate associations. We have called the nearer group Vel OB2A and the further group Vel OB2B.

De Zeeuw et al. list 41 candidates for Per OB2. These are widely dispersed in position and velocity space (radial velocity dispersion is 20 km $s^{-1}$). Our test converged to essentially the same group of candidates as for IC 348, and we have not listed Per OB2 separately in the data file. De Zeeuw et al. list 96 candidates for Lac OB1, also with a wide dispersion in position and velocity. Our list contains 68 members in a region with less than 6% of the 6-volume and a distance of 466 ± 17 pc. 31 of these have known radial velocity and 47 are contained in de Zeeuw's list. This group has a substantial overlap with ASCC 122. De Zeeuw et al. list 71 candidates for Cep OB2. Our list contains 13 members in a region with less than 0.4% of the 6-volume, and at a distance 882 ± 85 pc. 12 have known radial velocity and 7 are given in de Zeeuw's list. There is a substantial overlap between this group and NGC 7160. De Zeeuw et al. list 20 candidates for Cep OB6. Our list contains 11 members in a region with 18% of the 6-volume, and at a distance 275 ± 6 pc and with separation *n* = 4.5. 9 have known radial velocity and 10 are given in de Zeeuw's list.

### 5.5. Confirmed groups

Baumgardt (1998) reported the existence of Collinder 132, Roslund 5, and NGC 2451B as uncertain. We found 8 Hipparcos candidates for Collinder 132 with separation *n* = 5.5, whereas Baumgardt gave 12 candidates, none of which appear in our list. We found 18 Hipparcos stars in Roslund 5 with separation *n* = 3.5, compared to 5 Hipparcos stars given by Baumgardt, of which four are on our list. We identified 19 candidates for NGC 2451B with separation is *n* = 3.5. We thus confirm that these groups exist, and that NGC 2451B and Roslund 5 have characteristics of an association while Collinder 132 is a cluster.

### 5.6. Unidentifiable groups

In general it is not the case that one can exclude the existence of a moving group because no evidence for that group is found in Hipparcos. Exceptions arise for stellar associations in the Pleiades stream, because these are groups of young stars which ought to contain early types and should be represented in Hipparcos, and for certain lists given in the literature which are sufficiently dispersed in velocity and position space that they cannot be considered as moving groups.

Because the majority of stars participate in spiral structure, and therefore belong to streams, chance alignments are far more probable than would be the case for a well-mixed distribution, which has generally been assumed in the absence of a convincing explanation of spiral structure. It transpires that stars given as candidates for the "HIP 98321" group by Madsen, Dravins, and Lindegren (2002) are spread over a wide region of space and have a wide velocity dispersion. We concluded that they are not representative of a cluster or an association, but are merely stars in the Pleiades stream.

Similarly, the "HR 1614" group (Eggen, 1978) is not a moving group as such, but is a collection of stars in the Hercules stream. Similarities in stellar composition in HR 1614 reflect the known similarities in composition of the Hercules stream, which has been shown to consist predominately of stars predating the formation of spiral structure, about 9 Gyrs ago (Francis & Anderson, 2009).

We found no evidence for the TW Hydrae association or the η Chamaeleontis cluster which contain few Hipparcos stars with known radial velocities in the Pleiades stream. If these were true near associations they would be expected to contain early type stars from recent star formation and should be better represented in Hipparcos.

Of six stars listed by van Leeuwen (2009 & personal correspondence) as candidates for NGC 6633, five have known radial velocities, but these have widely discrepant values, with standard deviation, 17.3 km $s^{-1}$. It is impossible to say that these candidates are representative of a group.

According to Karchenko et al. and Dias et al., Collinder 121 is at a distance of 1100 pc — beyond the range of our tests. De Zeeuw et al. list 103 stars for Collinder 121. These have a mean parallax distance of 568 pc, but are widely dispersed in both velocity and position. We did not find any signs that they contain a moving group. The list given by de Zeeuw et al for Cas-Tau contains 83 stars. These are spread over a huge region of space. It is not possible to identify any core group or common motion by initialising a search from this list.

## 6. The Pleiades distance anomaly

The distance to the Pleiades has been calculated by many methods (e.g., Percival, Salaris & Groenewegen, 2005; Zwahlen et al., 2004; Turner, 1979; Soderblom et al., 2005; Xiaopei Pan, Shao & Kulkarni, 2004). The majority of determinations have found results in the range 132-135 pc (see van Leeuwen, 2009, table 3), but the Hipparcos satellite caused consternation by finding a distance of only 118 parsecs by measuring the parallax of stars in the cluster, a method which might be expected to yield the most direct and accurate result. In 2009, Van Leeuwen gave the revised distance, 120.2 ± 1.9 pc, using HIP2.

For the extended sample of 70 stars the mean parallax distance is 124.1 ± 2.7 pc. The major part of the increase results because we calculated the distance from the mean parallax without weighting, while Van Leeuwen found mean parallax weighted by inverse error squared. Because parallax distances to multiple star systems are inherently less accurate, and may contain an unquantified systematic error, a probably better estimate is obtained by considering only single stars. From 45 single stars we obtain a distance of 125.6 ± 4.2 pc. This is typically within 2$\sigma$ of the total error of the value found by other methods.

Van Leeuwen (2009) suggested that unknown factors in stellar evolution may result in an unmodelled difference between the positions of the Pleiades and Hyades on the CMD, and that using nearby stars of similar age and composition to the Pleiades might improve the calculation of the Pleiades distance from luminosity. In fact there is a small difference between the position of the Hyades on the CMD and the position of other groups. This is expected since the Hyades has greater metalicity and age. AB Doradus, $\beta$ Pictoris, and Tucana/Horologium (and in particular AB Doradus) are thought to be similar in age and composition to the Pleiades. Indeed, AB Doradus has very close motion, and is thought to be coeval and to have an origin in space very close to the Pleiades (Ortega et al. 2007). Based on these three near groups we find photometric distance, 132 ± 3 pc, slightly less than estimates based on the Hyades. We conclude that the Hipparcos parallax distance for the Pleiades is low, but consistent within measurement errors.

Van Leeuwen grouped the Diamond cluster (NGC 2516) together with the Pleiades and Blanco 1 as having anomalously low parallax distances. For the Diamond cluster, Van Leeuwen used 11 stars, of which we have rejected three, while our list contains 21 stars, 19 with radial velocities. For our complete list of 21 stars, we found distance 388 ± 22 pc, and for singles stars we found distance 395 ± 24 pc. These figures are in good agreement with photometric distances to the Diamond cluster given by van Leeuwen. The difference in membership accounts for much of the anomaly reported by Van Leeuwen; we calculate 360 ± 14 pc, compared to 342.5 pc calculated by van Leeuwen for the same stars.

We found 16 candidates for Blanco 1, 15 with radial velocities, with separation $n$ = 6. Van Leeuwen found 13 candidates, of which we reject one. For our list of 16 stars we found a distance of 205 ± 15 pc. For 13 single stars we found 214 ± 14 pc, compared to 207 ± 12 pc given by van Leeuwen. Hipparcos stars in Blanco 1 are too early to obtain a reliable photometric distance, but Stauffer et al. (2010) provide data for a larger sample of less bright candidate single stars, from which we find photometric distance 235 ± 10 pc. This discrepancy between photometric and parallax distances is not excessive.

Comparison between the calculated parallax distances from XHIP and distances given by Dias et al. and Karchenko et al. shows a small number of other discrepancies, which may be due to differences in candidate lists, but no systematic difference between photometric and parallax distance.

## 7. Conclusion

By making use of up to date radial velocity information to fit 3D velocity as well as position, we have found improved lists of candidates in the Hipparcos data for 87 clusters and associations. We have defined a parameter $n$, for separation from surrounding field stars, and used it to divide these into 42 likely clusters and 45 associations. We have calculated size and parallax distance and size for these moving groups.

We observed that the calculation of the distance to the Pleiades from Hipparcos data by van Leeuwen contains a subtle systematic error, which biases the calculation towards a low result, and that parallax distances and luminosities of binaries and multiple star systems are inherently less accurate than those of single stars. For our sample of 45 single Pleiades stars we found a mean parallax distance 125.6 ± 4.2 pc. This result is less than 2$\sigma$ below most estimates by other methods, and 1.2$\sigma$ less than the photometric distance, 132 ± 3 pc, calibrated to AB Doradus, $\beta$ Pictoris, and Tucana/Horologium, which are thought to be similar in age and composition. The Pleiades parallax distance is thus no longer a cause for concern.

**Data**

XHIP can be retrieved from the Centre de Données astronomiques de Strasbourg (CDS Catalog V/137B). The list of found groups and their properties is contained in the file groups.dat.